\documentclass[10pt,journal,cspaper,compsoc]{IEEEtran}
\ifCLASSOPTIONcompsoc
\else
\fi
\ifCLASSINFOpdf
   \usepackage[pdftex]{graphicx}
   \DeclareGraphicsExtensions{.eps,.pdf,.jpeg,.png}
\else
\fi
\usepackage[cmex10]{amsmath}
\usepackage{amssymb} 
\usepackage{url}

\usepackage{multirow, array, framed, fixltx2e} 
\usepackage{epstopdf}

\begin{document}
%
\title{Solid State Disk Object-Based Storage\\with Trim Commands}
%
%
%
%

\author{Tasha~Frankie,~\IEEEmembership{Member,~IEEE,}
        Gordon~Hughes,~\IEEEmembership{Fellow,~IEEE,}
        and~Ken~Kreutz-Delgado,~\IEEEmembership{Senior Member,~IEEE}
\IEEEcompsocitemizethanks{\IEEEcompsocthanksitem T. Frankie and K. Kreutz-Delgado are with the Department of Electrical and Computer Engineering, UC San Diego, La Jolla, 92093.\protect\\
E-mail: tcvanesi@ucsd.edu, kreutz@eng.ucsd.edu
\IEEEcompsocthanksitem G. Hughes is with the Center for Magnetic Recording Research, UC San Diego, La Jolla, 92093.\protect\\
E-mail: ghughes@alumni.caltech.edu}
\thanks{}}

\IEEEcompsoctitleabstractindextext{%
\begin{abstract}
This paper presents a model of NAND flash SSD utilization and write amplification when the ATA/ATAPI SSD Trim command is incorporated into object-based storage under a variety of user workloads, including a uniform random workload with objects of fixed size and a uniform random workload with objects of varying sizes.  We first summarize the existing models for write amplification in SSDs for workloads with and without the Trim command, then propose an alteration of the models that utilizes a framework of object-based storage.  The utilization of objects and pages in the SSD is derived, with the analytic results compared to simulation.  Finally, the effect of objects on write amplification and its computation is discussed along with a potential application to optimization of SSD usage through object storage metadata servers that allocate object classes of distinct object size.
\end{abstract}

\begin{keywords}
Trim, Solid State Drive, Object-Based Storage, Write Amplification, NAND Flash, Markov processes.
\end{keywords}}

\maketitle

\IEEEdisplaynotcompsoctitleabstractindextext

%
\IEEEpeerreviewmaketitle

\section{Introduction}
%
%

%
%
%
%
\IEEEPARstart{I}{n} recent years, NAND flash solid state drives (SSDs) have become ubiquitous, appearing in small mobile electronics to large data centers.  Part of the popularity of flash-based SSDs over traditional hard disk drives (HDDs) is the increased speed and performance that they offer.  However, SSDs have their own set of problems, including a slowdown in performance as the device fills with data due to write amplification, and a limited lifetime due to wearout of the floating-gate transistors.  Additionally, SSDs currently need to conform with legacy interfaces from HDDs, which makes optimization in using the device more challenging.

Object based storage offers multiple advantages for systems using solid state storage devices~\cite{kang2010efficient}. When the multiple thousand pages in a flash block are divided into object records of larger size, the resultant fewer objects per block directly translates into fewer data write relocations during garbage collection, leading to a lower write amplification and faster performance. This offers an opportunity for user application storage metadata, via metadata servers, to actively allow flash storage devices to self-optimize their hardware, to fully utilize SSD write speed advantages over magnetic HDD while ameliorating SSD limitations like block erase and finite write lifetime. Object storage metadata servers could allocate SSDs to object classes~\cite{feng2006adaptive} of distinct object size and thereby allow SSD to internally manage their object physical mapping and relocation. This has the potential to improve I/O rate and QoS, user access blocking probability, and SSD flash chip lifetime through the reduction of write amplification.

In this paper, we present a model for computing utilization and write amplification under a modified random workload including Write and Trim commands, and data records of varying sizes.  This model is an extension of models proposed for simpler workloads by previous researchers~\cite{Hu2009,hu2010fundamental,agarwal2010closed,xiang2011improved} and builds on our previous models of the Trim command~\cite{frankie2012_acmse,frankie2012_ciit,frankie2012TOS}. Good models assist in the development of algorithms for managing the challenges of SSDs; our model adds flexibility to the theory in order to more closely match real-world workload characteristics.

%

\section{NAND Flash SSDs}
In this section, we provide background about NAND Flash SSDs which is necessary to understand the model presented in this paper.

\subsection{Layout}
The organization of a NAND flash SSD gives it characteristics that cause unique challenges.  The smallest storage unit of flash is the \emph{page}, which typically holds 2K, 4K, or 8K of data~\cite{grupp2009characterizing,roberts2009integrating}.  Pages are grouped into \emph{blocks}, with blocks typically containing 64, 128, or 256 pages, varying by manufacturer~\cite{grupp2009characterizing}.  One challenge of flash SSDs is that a page must be erased before it can be programmed again, but the smallest erase unit of flash SSDs is the block, and not all pages on the block are ready to be erased at the same time.  This means that write-in-place schemes are impractical, and most SSDs employ a dynamic logical-to-physical mapping called the flash translation layer (FTL)~\cite{chung2009survey,gupta2009dftl}.  With this dynamic mapping, when data is overwritten, the old physical location of the data is marked as invalidated in the FTL mapping.  In this paper, we utilize a log-structured file system, in which data is written to successively available pages, as described in~\cite{Hu2009}.

\subsection{Garbage Collection}
Garbage collection is the process of selecting and erasing a block to reclaim space for writing.  When a block is chosen for erasure, any valid pages on the block first need to be copied elsewhere to avoid data loss.  Copying these valid pages leads to write amplification, a phenomenon in which the device performs more writes than are requested of it.

There are many algorithms used to select a block for garbage collection, most of which are based on one of two schemes originally described in the seminal paper by Rosenblum and Ousterhout~\cite{rosenblum1992design}.  Their first method of selection chooses the block that creates the most free space, and their second method also accounts for the age of the block.  In this paper, we utilize greedy garbage collection, in which the block with the fewest valid pages is selected.

\subsection{Improving Performance}
Reducing write amplification creates an improvement in performance by increasing the average write speed due to the fewer number of extra writes performed by the device.  Additionally, by reducing the number of program/erase cycles, SSD lifetime is extended, since flash devices wear out and lose the ability to store data after being written many times.

There are several ways to reduce write amplification.  One common way is for manufacturers to provide extra storage space on the SSD beyond what the user is allowed to access, a practice called overprovisioning.  Overprovisioning reduces write amplification because the amount of time between garbage collections increases, which tends to lead to a decrease in the number of valid pages in the block selected for reclamation.  We measure the amount of overprovisioning with the spare factor, $S_f=(T-u)/T, 0 \le S_f \le 1$, where $T$ is the total number of pages on the SSD, and $u$ is the number of pages in the user space.

Another way to reduce write amplification is through use of the Trim command~\cite{TrimSpec}, which is a way for the file system to communicate to the SSD that certain data pages no longer need to be retained.  The pages associated with the data are invalidated in the FTL mapping, meaning that there are fewer valid pages that need copying during garbage collection.

\section{Analyzing Performance}\label{sec:previous_work}
Several researchers have created analytic models to help understand the performance of SSDs.  In this section, we summarize previous work that we build on in this paper.

\subsection{Uniform Random Workload}
The type of workload used can have a large effect on the performance of a storage device~\cite{agrawal2008design}.  We view real-world workloads as existing along a continuum from purely sequential workloads to completely random workloads.  For SSDs, analysis of a purely sequential workload is uninteresting, because entire blocks of data are invalidated before garbage collection, leading to no write amplification~\cite{Hu2009}.  However, at the other end of the spectrum, completely random workloads are very interesting to statistically analyze, since the random nature of the writes results in most blocks still containing valid pages at the time of garbage collection, leading to potentially high write amplification.
Hu, et al.~\cite{Hu2009,hu2010fundamental}, Agarwal and Marrow~\cite{agarwal2010closed}, and Xiang and Kurkoski~\cite{xiang2011improved} have all proposed models for write amplification of the uniform random workload under greedy garbage collection.  Their workloads consist of write requests distributed uniformly randomly over the user space, with each logical block address (LBA) requested requiring one page to store the data.  Each of their models found that the single most important factor in determining the write amplification was the level of overprovisioning provided on the SSD.

\subsection{Trim-Modified Uniform Random Workload}
Previously, we modified the uniform random workload to incorporate Trim requests, then analyzed the resulting utilization and write amplification~\cite{frankie2012_acmse,frankie2012_ciit,frankie2012TOS}.  In the Trim-modified uniform random workload, both write and Trim requests are allowed, and the data associated with each LBA occupies one page.  As in the standard uniform random workload, write requests are uniformly randomly chosen over all $u$ user LBAs, and Trim requests are uniformly randomly chosen over all In-Use LBAs, where an LBA is defined as being In-Use if its most recent request was a Write.  Trim requests occur with probability $q$, and write requests occur with probability $1-q$.  This workload was modeled as a Markov birth-death chain with the state being the number of In-Use LBAs, and the steady state solution was approximated as Gaussian with mean $us$ and variance $u\bar{s}$, where $s=(1-2q)/(1-q)$ and $\bar{s}=q/(1-q)$ are dependent on the rate of Trim.

\subsubsection{Effective Overprovisioning}
Since the number of In-Use LBAs is the same as the number of valid pages, the level of effective overprovisioning is straightforward to calculate.  The effective overprovisioning is measured in terms of the effective spare factor, $S_\text{eff} = (T-X_n)/T$, where $X_n$ is the number of valid pages at time $n$.  The mean effective spare factor was found to be
\begin{eqnarray*}
\bar{S}_{\text{ eff}} &=& \frac{T-us}{T}\\
&=& \bar s + s S_f
\end{eqnarray*}
The variance of the effective spare factor was also calculated, and found to be negligible in practical devices (as $T \to \infty$)~\cite{frankie2012_acmse,frankie2012TOS}.

\subsubsection{Write Amplification}
We also explored write amplification under the Trim-modified random workload.  After paralleling the derivation of Xiang and Kurkoski~\cite{xiang2011improved}, we discovered that all three previously proposed models for write amplification could be modified by using the effective level of overprovisioning.  In simulation, the Trim-modified model of Xiang and Kurkoski provided the best prediction; the Trim-modified models of Hu, et al., and of Agarwal and Marrow were optimistic in their predictions of write amplification~\cite{frankie2012_ciit,frankie2012TOS}.

Additionally, we separated the workload into hot (frequently accessed) and cold (infrequently accessed) data in order to closer approximate a real-work workload.  Analysis of this workload led to a more general workload with $N$ temperatures, and found that the write amplification of this workload type was lowest when the $N$ temperatures were kept physically separated into different blocks~\cite{frankie2012TOS}.

\section{Model Augmentation}
In the previous work described in Section~\ref{sec:previous_work}, one page stored the data associated with one LBA. In this section, we augment the model to allow larger amounts of data to be stored under a single identifier, which lends itself naturally to a discussion of object-based storage.

\subsection{From LBAs to Objects}
In object-based storage, the filesystem does not assign data to particular LBAs.  Instead, the filesystem tells the object-based storage disk to store certain data; the disk chooses where to store it, and gives the filesystem a unique object ID with which the data can be retrieved and/or modified~\cite{OSDSpec}.  Object-based storage allows for variable-sized objects with identifiers that need not indicate any information about the physical location of the data~\cite{kang2011object}.  Rajimwale, et al. have proposed that object-based storage is an optimal interface for flash-based SSDs~\cite{rajimwale2009block}.  We use object-based storage as motivation for extensions to the theory of effective overprovisioning and write amplification described in Section~\ref{sec:previous_work}.

The utilization problem we solved previously~\cite{frankie2012_acmse} can be viewed in object space instead of LBA space.  Instead of LBAs that are In-Use or not In-Use, there are object IDs that are In-Use or not In-Use.  Previously, there were $u$ LBAs allowed; now we allow object IDs in the range $[1,u]$, for a maximum of $u$ objects.  The object-view of the Trim-modified uniform random workload is the same as the LBA view of the problem when each object occupies 1 page.

To remove the restriction that one object occupies only one page, we start by examining the workload in which each object occupies $b$ pages, where $b$ is constant and identical for all objects in the workload.  Then, we  examine the workload in which $b$ varies each time an object is written.  For all workloads, we assume that when an object ID is selected for trimming (or, more appropriately for objects, deletion), all pages associated with that object ID are invalidated.  Likewise, when an object is rewritten, all pages previously containing data associated with that object ID are invalidated.  Note that there are now two quantities of interest: the number of object IDs that are in-use, and the number of pages that are valid.  When each object occupies one page, these two quantities are the same.

Define the following variables:

$\chi_t = (\chi_{1,t}, \chi_{2,t}, \ldots, \chi_{u,t})^\mathrm{T}$ is a vector of indicator random variables at time $t$.

\begin{equation*}
\chi_{i,t} = \textbf{1}_{A_i}(\omega) = \begin{cases}1& \omega \in A_i \text{ (object i is In-Use)}\\
0& \text{otherwise (object i is not In-Use)}\end{cases}
\end{equation*}
where
{\small
\begin{align*}
A_i =
 \{\omega | \text{the STATE}(\omega) \text{ is such that location $i$ is occupied}\}
\end{align*}
}

$X_t=\sum_{i=1}^u \chi_{i,t}$ is a random variable representing the number of objects in-use at time $t$.  The mean and variance of this random variable at steady state were computed previously by Frankie, et al.~\cite{frankie2012_acmse,frankie2012TOS}.

$Z_t=(Z_{1,t}, Z_{2,t}, \ldots, Z_{u,t})^\mathrm{T}$ is a vector of the size of each object at time $t$.

Now, compute the number of valid pages at time $t$: $Y_t=Z_t^\mathrm{T}\chi_t = \sum_{i=1}^u \chi_{i,t}Z_{i,t}$

We are concerned with the values at steady state, and drop the time index to indicate the steady state solution.

\subsection{Fixed Sized Objects}
This workload is an extension of the original workload whose steady state pdf was derived in~\cite{frankie2012_acmse}.  However, objects now occupy $b$ pages instead of 1 page.  Recall that the number of In-Use objects is the same as the number of In-Use LBAs previously derived; only the number of valid pages will change.  We assume that at least $ub$ pages are allowed for the user to write to, so that it is not possible to request more space than is available.

\medskip
For this workload, $Z_t=(b, b, \ldots, b)^\mathrm{T} \; \forall t$. This means
\begin{eqnarray*}\label{}
Y_t &=& \sum_{i=1}^u \chi_{i,t}Z_{i,t} \\
    &=& \sum_{i=1}^u \chi_{i,t}b \\
    &=& b \sum_{i=1}^u \chi_{i,t} \\
    &=& b X_t
\end{eqnarray*}
To compute the steady state mean and variance of $Y$, we can use the known steady state mean and variance of $X$.
\begin{eqnarray*}
E[Y] &=& E[bX] \\
     &=& b E[X] \\
     &=& bus
\end{eqnarray*}
Similarly,
\begin{eqnarray*}
Var[Y] &=& Var[bX] \\
     &=& b^2 Var[X] \\
     &=& b^2u\bar{s}
\end{eqnarray*}

\subsection{Variable Sized Objects}
In this workload, the size of each object varies independently over time.  Each time an object ID is selected to be written, the size of the object is chosen randomly, according to the distribution from which $Z_{i,t}$ is drawn, regardless of any size that object ID may have had in the past.  $Z_{i,t}$ is drawn from a distribution with a minimum value $B_1 \ge 1$, a maximum value $B_2 \ge B_1$, a mean value $m_Z$, and a standard deviation $\sigma_Z$.  Again, we assume there is sufficient space for writing all requests; this means there are at least $uB_2$ pages allowed for the user.  Because $Z_{i,t}$ is not constant, $Y_t = \sum_{i=1}^u \chi_{i,t}Z_{i,t}$ cannot be simplified, as it was in the fixed-sized object case.

The mean is still relatively straightforward to calculate, since $\chi_i$ and $Z_i$ are independent, and the $Z_i$s are iid:
\begin{eqnarray}
E[Y] &=& E\left[\sum_{i=1}^u \chi_{i}Z_{i}\right] \nonumber\\
     &=& \sum_{i=1}^u E[\chi_{i}Z_{i}] \nonumber\\
     &=& \sum_{i=1}^u E[\chi_{i}]E[Z_{i}] \nonumber\\
     &=& \sum_{i=1}^u E[\chi_{i}]m_Z \nonumber\\
     &=& m_Z\sum_{i=1}^u E[\chi_{i}] \nonumber\\
     &=& m_Z E\left[\sum_{i=1}^u\chi_{i}\right] \nonumber\\
     &=& m_Z E[X] \nonumber\\
     &=& m_Z u s \label{eqn:var_obj_mean}
\end{eqnarray}

The variance is more complicated because of cross-terms.  The $\chi_i$ values are not independent, so it is not simple to compute the variance\footnote{If the $\chi_i$ values were independent, $X$ would have a binomial distribution, but because we know from previous analysis that the variance is $u\bar{s}$ and not $us\bar{s}$, this is obviously not the case.}.

To compute the variance of $Y$, compute
\begin{equation*}
Var(Y) = E[Y^2] - E[Y]^2
\end{equation*}
The value of $E[Y]$ was computed in Equation~\ref{eqn:var_obj_mean}.  Computing $E[Y^2]$ is more complicated.  When following the steps below, note that $\chi_i^2 = \chi_i$ , $\chi_i$ is independent of $Z_i$ and $Z_j$, and the $Z_i$ values are iid.
\begin{eqnarray*}
E[Y^2] &=& E\left[ \left(\sum_{i=1}^u \chi_i Z_i \right)^2 \right] \\
       &=& E\left[ \sum_{i=1}^u \chi_i Z_i^2 + 2 \sum_{i<j} \chi_i \chi_j Z_i Z_j \right] \\
       &=& \sum_{i=1}^u E\left[\chi_i Z_i^2 \right] + 2 \sum_{i<j} E\left[\chi_i \chi_j Z_i Z_j \right] \\
       &=& \sum_{i=1}^u E\left[\chi_i \right] E\left[Z_i^2 \right] \\
       && + 2 \sum_{i<j} E\left[\chi_i \chi_j \right] E\left[Z_i \right] E\left[Z_j \right] \\
\end{eqnarray*}
Most of these expected values are straightforward to calculate.  However, care must be taken when computing $E[\chi_i \chi_j]$, because the $\chi_i$ values are NOT independent.  See Section~\ref{sec:appendix} for details.

Continuing,
\begin{eqnarray*}
E[Y^2] &=& (\sigma_Z^2 + m_Z^2) us + 2 m_Z^2 \sum_{i<j} \frac{s \left(us- \left(1-\frac{\bar{s}}{s} \right) \right)}{u-1}\\
       &=& (\sigma_Z^2 + m_Z^2) us + 2 m_Z^2 \frac{(u-1)u}{2} \frac{s \left(us- 1+\frac{\bar{s}}{s}  \right)}{u-1} \\
       &=& \sigma_Z^2 us + m_Z^2 us + m_Z^2 us \left( us - 1 + \frac{\bar{s}}{s}  \right) \\
       &=& \sigma_Z^2 us + (m_Z us)^2 + m_Z^2 u\bar{s}
\end{eqnarray*}

Now it is possible to compute
\begin{eqnarray*}
Var(Y) &=& E[Y^2] - E[Y]^2  \\
       &=& \sigma_Z^2 us + (m_Z us)^2 + m_Z^2 u\bar{s} - (m_Z us)^2 \\
       &=& \sigma_Z^2 us + m_Z^2 u\bar{s}
\end{eqnarray*}

\subsubsection{Computing $E[\chi_i \chi_j]$} \label{sec:appendix}
As mentioned previously, care must be taken when computing $E[\chi_i \chi_j]$, because the $\chi_i$ values are NOT independent.
\begin{eqnarray*}
E[\chi_i \chi_j] &=& P(A_i \cap A_j)  \\
                 &=& P(A_i|A_j)P(A_j)
\end{eqnarray*}
Obviously, when $i=j$, $P(A_i|A_j)=1$.  However, calculating this value when $i\neq j$ is important.  Since there is nothing special about any particular $i$ or $j$, assume that $P(A_i|A_j)$ is the same for all $i\neq j$, and thus is a constant with respect to $i$ and $j$.

It is tempting to claim that $P(A_i|A_j) = (us-1)/(u-1)$, since $us$ is the expected number of In-Use objects at steady state.  However, this is an invalid statement because it implicitly assumes a zero variance.  In fact, using this value in the calculation of $Var(Y)$ results in a statement that $Var(Y) = 0$, which is exactly what is expected because of the implicit zero variance assumption.

Instead of trying to compute $P(A_i|A_j)$ directly, it is possible to compute it by using the result of the one object occupies one page model.  Working backwards to calculate this probability, and noting that there are $\frac{(u-1)u}{2}$ terms in the sum over $i<j$:
\begin{eqnarray*}
Var(Y) &=& Var\left(\sum_{i=1}^u \chi_i\right) \\
       &=& \sum_{i=1}^u Var\left(\chi_i\right) + 2 \sum_{i<j} Cov(\chi_i, \chi_j) \\
       &=& \sum_{i=1}^u s \bar{s} + 2 \sum_{i<j} C \\
u\bar{s}&=&us\bar{s} + 2u(u-1) C
\end{eqnarray*}
Solving for $C$ yields $\frac{\bar{s}^2}{u-1}$, which is the value of the covariance.\footnote{Interestingly, as $u$ gets large, the covariance goes to 0, indicating that the $\chi_i$ values are effectively independent for large $u$.  However, the rate at which the covariance goes to 0 is counteracted by the rate at which the number of terms in the sum over $i<j$ increases, resulting in a variance that is NOT binomially distributed.}  Continuing with the calculations,
\begin{eqnarray*}
Cov(\chi_i, \chi_j) &=& C \\
                    &=& E[\chi_i \chi_j] - P(A_i)P(A_j)\\
                    &=& E[\chi_i \chi_j] - s^2\\
\end{eqnarray*}
so that
\begin{eqnarray}\label{eqn:conditional_wanted}
E[\chi_i \chi_j] &=& C + s^2 \nonumber \\
                 &=& \frac{\bar{s}^2}{u-1} + s^2 \nonumber \\
                 &=& \frac{s \left(us- 1 + \frac{\bar{s}}{s} \right)}{u-1}
\end{eqnarray}

\subsection{Simulation Results}

\begin{figure}[!t]
\centering
\includegraphics[width=3.0in]{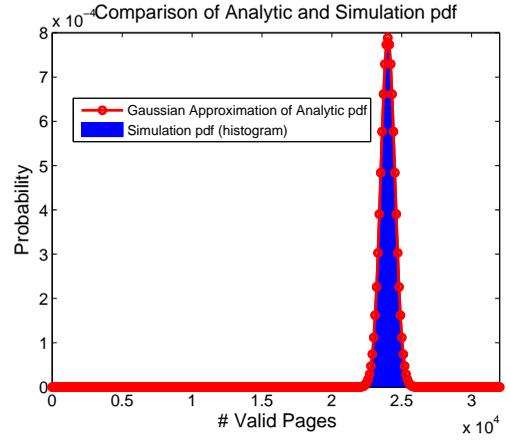}
\caption{Comparison of analytic and simulation pdf for the number of valid pages when objects are a fixed size of 32 pages.  For this example, $u=1000$ and $q=0.2$.}
\label{fig:fixed_size_pdf}
\end{figure}

\begin{table*}
\centering
\caption{Mean and $\sigma$ of Number of In-Use Objects and Number of Valid Pages for Fixed Sized Objects.  Values in this table were computed with $u=1000$.  The size of the objects was fixed at 32. Simulation values are the average of 64 Monte-Carlo simulations.}
\begin{tabular}{*{9}{|c}|} \hline
\multirow{2}{*}{Trim Amount $q$}&\multicolumn{2}{|c|}{Mean Objects} & \multicolumn{2}{|c|}{$\sigma$ Objects} & \multicolumn{2}{|c|}{Mean Pages} & \multicolumn{2}{|c|}{$\sigma$ Pages}\\ \cline{2-9}
    & Analytic & Simulation & Analytic & Simulation & Analytic & Simulation & Analytic & Simulation\\ \hline
0.05 & 947.37 & 947.37 & 7.25 & 7.24 & 30315.79 & 30315.80 & 232.15 & 231.80\\ \hline
0.1 & 888.89 & 888.89 & 10.54 & 10.54 & 28444.44 & 28444.37 & 337.31 & 337.37\\ \hline
0.2 & 750.00 & 750.01 & 15.81 & 15.82 & 24000.00 & 24000.37 & 505.96 & 506.18\\ \hline
0.3 & 571.43 & 571.47 & 20.70 & 20.73 & 18285.71 & 18287.10 & 662.46 & 663.29\\ \hline
0.4 & 333.33 & 333.43 & 25.82 & 25.83 & 10666.67 & 10669.70 & 826.24 & 826.46\\ \hline
0.45 & 181.82 & 181.89 & 28.60 & 28.62 & 5818.18 & 5820.40 & 915.32 & 915.93\\
\hline\end{tabular}
\label{table:fixed_sized_objs}
\end{table*}

Monte-Carlo simulation was used to verify the analytic results, with various values of $u$, $q$, and distributions of $Z$.  In this paper, results are shown for $u=1000$.  Simulations were run for 1 million requests to ensure steady state was reached, then the simulation means and variances were averaged over the next 9 million requests.  The results in each table are the average of 64 Monte-Carlo simulations.  Although results are only shown for $u=1000$, results for larger values of $u$ are comparable.

Table~\ref{table:fixed_sized_objs} shows the analytic and simulation values of the mean and standard deviation of the number of In-Use objects as well as the mean and standard deviation of the number of valid pages at steady state when the objects are all a fixed size of 32 pages.  When rounded to the nearest integer, the mean number of objects is predicted exactly by analysis; the mean number of valid pages is predicted to within 3 pages.  The slight error in the prediction of the number of valid pages is likely due to the extremely large variances seen when $q\ge0.3$. An overlay of the simulation pdf and the analytic pdf for the number of valid pages is shown in Fig.~\ref{fig:fixed_size_pdf} for $q=0.2$.

\begin{figure}[!t]
\centering
\includegraphics[width=3.0in]{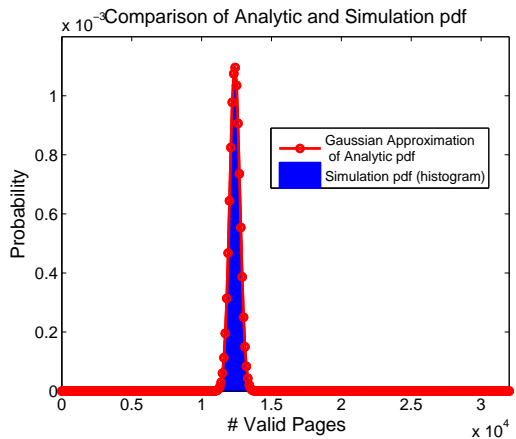}
\caption{Comparison of analytic and simulation pdf for the number of valid pages when the size of objects is drawn from a discrete uniform distribution over $[1,32]$.  For this example, $u=1000$ and $q=0.2$.}
\label{fig:unif_dist_pdf}
\end{figure}

Table~\ref{table:variable_sized_objs} shows the analytic and simulation values of the mean and standard deviation of the number of In-Use objects as well as the mean and standard deviation of the number of valid pages at steady state when the size of the objects is drawn from a discrete uniform distribution over $[1,32]$.  Analytic and simulation values match exactly when rounded to the nearest integer.  An overlay of the simulation pdf and the analytic pdf for the number of valid pages is shown in Fig.~\ref{fig:unif_dist_pdf} for $q=0.2$.

\begin{table*}
\centering
\caption{Mean and $\sigma$ of Number of In-Use Objects and Number of Valid Pages for Variable Sized Objects.  Values in this table were computed with $u=1000$.  The size of the objects was drawn from a discrete uniform distribution over $[1,32]$. Simulation values are the average of 64 Monte-Carlo simulations.}
\begin{tabular}{*{9}{|c}|} \hline
\multirow{2}{*}{Trim Amount $q$}&\multicolumn{2}{|c|}{Mean Objects} & \multicolumn{2}{|c|}{$\sigma$ Objects} & \multicolumn{2}{|c|}{Mean Pages} & \multicolumn{2}{|c|}{$\sigma$ Pages}\\ \cline{2-9}
    & Analytic & Simulation & Analytic & Simulation & Analytic & Simulation & Analytic & Simulation\\ \hline
0.05 & 947.37 & 947.36 & 7.25 & 7.27 & 15631.58 & 15631.33 & 308.37 & 307.92\\ \hline
0.1 & 888.89 & 888.92 & 10.54 & 10.53 & 14666.67 & 14667.03 & 325.62 & 325.63\\ \hline
0.2 & 750.00 & 750.00 & 15.81 & 15.80 & 12375.00 & 12374.67 & 363.32 & 363.42\\ \hline
0.3 & 571.43 & 571.41 & 20.70 & 20.70 & 9428.57 & 9429.00 & 406.69 & 406.49\\ \hline
0.4 & 333.33 & 333.35 & 25.82 & 25.79 & 5500.00 & 5500.13 & 458.17 & 457.71\\ \hline
0.45 & 181.82 & 181.81 & 28.60 & 28.58 & 3000.00 & 2999.85 & 488.11 & 487.65 \\
\hline\end{tabular}
\label{table:variable_sized_objs}
\end{table*}

\begin{figure}[!t]
\centering
\includegraphics[width=3.0in]{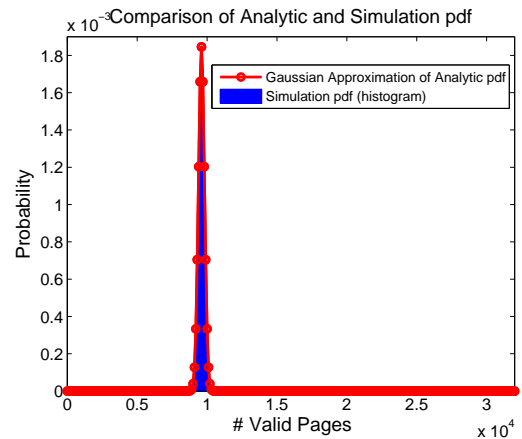}
\caption{Comparison of analytic and simulation pdf for the number of valid pages when the size of objects is drawn from a binomial distribution over with parameters $p=0.4$ and $n=32$.  For this example, $u=1000$ and $q=0.2$.}
\label{fig:bino_dist_pdf}
\end{figure}

Table~\ref{table:variable_sized_objs_bino} shows the analytic and simulation values of the mean and standard deviation of the number of In-Use objects as well as the mean and standard deviation of the number of valid pages at steady state when the size of the objects is drawn from a binomial distribution with parameters $p=0.4$ and $n=32$.  Analytic and simulation values match exactly when rounded to the nearest integer.  An overlay of the simulation pdf and the analytic pdf for the number of valid pages is shown in Fig.~\ref{fig:bino_dist_pdf} for $q=0.2$.

\begin{table*}
\centering
\caption{Mean and $\sigma$ of Number of In-Use Objects and Number of Valid Pages for Variable Sized Objects.  Values in this table were computed with $u=1000$.  The size of the objects was drawn from a binomial distribution over with parameters $p=0.4$ and $n=32$.  Simulation values are the average of 64 Monte-Carlo simulations.}
\begin{tabular}{*{9}{|c}|} \hline
\multirow{2}{*}{Trim Amount $q$}&\multicolumn{2}{|c|}{Mean Objects} & \multicolumn{2}{|c|}{$\sigma$ Objects} & \multicolumn{2}{|c|}{Mean Pages} & \multicolumn{2}{|c|}{$\sigma$ Pages}\\ \cline{2-9}
    & Analytic & Simulation & Analytic & Simulation & Analytic & Simulation & Analytic & Simulation\\ \hline
0.05 & 947.37 & 947.36 & 7.25 & 7.27 & 12126.32 & 12126.39 & 126.09 & 126.06\\ \hline
0.1 & 888.89 & 888.88 & 10.54 & 10.53 & 11377.78 & 11377.73 & 158.21 & 158.10\\ \hline
0.2 & 750.00 & 750.01 & 15.81 & 15.80 & 9600.00 & 9600.25 & 216.15 & 216.00\\ \hline
0.3 & 571.43 & 571.37 & 20.70 & 20.68 & 7314.29 & 7313.46 & 273.14 & 272.77\\ \hline
0.4 & 333.33 & 333.32 & 25.82 & 25.79 & 4266.67 & 4266.57 & 334.35 & 334.03\\ \hline
0.45 & 181.82 & 181.84 & 28.60 & 28.63 & 2327.27 & 2327.60 & 368.03 & 368.36\\
\hline\end{tabular}
\label{table:variable_sized_objs_bino}
\end{table*}

\section{Application to Write Amplification}
In this section, we examine write amplification when one object requires a fixed number of pages to store data.  We assume that the number of pages required to store the data associated with the object evenly divides the number of pages per block (i.e. $mod(n_p, b) = 0$, so that an entire object is stored within a single block.  As in previous work, data is stored in the fashion of a log-structured file system with greedy garbage collection.

We argue that fixed sized objects with sizes that evenly divide the number of pages per block is equivalent to reducing the number of pages in a block.  For instance, if two objects can fit in one block, this is equivalent to having a block with two (large size) pages, and objects that require one of these (large) pages to store the associated data.  With this equivalence between reduced number of pages per block and increased object size, the formulas for write amplification with Trim that we previously developed~\cite{frankie2012_ciit} should apply.  However, the Trim-modified Agarwal and the Trim-modified Xiang models do not take $n_p$ into account!  The derivation of the Trim-modified Agarwal model causes $n_p$ to drop out naturally; it is not a good candidate for modification.  The Trim-modified Xiang model is a better candidate, as asymptotically removes $n_p$ in the derivation.  By eliminating this asymptotic approximation, the write amplification is computed as $n_p/y$, where
\footnotesize
\begin{align}
\text{\normalsize $y$} =& \text{\normalsize $\ n_p \  +$} \nonumber\\
  & \frac{\text{\small$-W$}\left(T (1+\bar{s}) \left(1-\text{\small$\frac{1}{s(1+\bar{s})u}$} \right)^{T(1+\bar{s})} \text{ln}\left(1-\text{\small$\frac{1}{s(1+\bar{s})u}$} \right) \right)} {\frac{T}{n_p}(1+\bar{s}) \text{ln}\left(1-\text{\small$\frac{1}{s(1+\bar{s})u}$} \right)} \label{eqn:modified_Xiang}
\end{align}
\normalsize
and $W(\cdot)$ is the Lambert W function~\cite{corless1996lambertw,xiang2011improved,frankie2012_ciit}.
The Trim-modified Hu model does not need to be changed in order to take $n_p$ into account, as it already does this.  The model can be found in Figure~\ref{fig:hu_model}.

\begin{figure}[t!]
\begin{framed}
Computation of Trim-Modified Write Amplification Factor
\medskip
\hrule \vspace{8pt}
Compute $p = (p_1, \cdots, p_j, \cdots, p_w)$
\begin{eqnarray}
p_1 &=& e^{\text{\normalsize$-1.9\left(\frac{T}{us}-1\right)$}} \label{eqn:Hu_needs_modify} \\
p_j &=& \text{min}\left(1, \; \frac{1.1 \, p_{j-1}}{ \left(1-\frac{1}{us}\right)^{n_p}} \right) \label{eqn:Hu_needs_modify2}
\end{eqnarray}
\hrule \vspace{8pt}
Compute $v = \big(v_{j,k}\big)$, a matrix of size $w \times n_p$, \newline \phantom{1} \hspace{10.2mm} $j= 1, \cdots, w$\quad \& \ \  $k= 0, \cdots, n_p-1$
\begin{equation*}
v_{j,k} = 1 - \sum_{m=0}^{k}\left[{n_p \choose m} p_j^m \left(1-p_j\right)^{n_p-m}\right]
\end{equation*}
\hrule \vspace{8pt}
Compute $V = (V_1, \cdots, V_k, \cdots, V_{n_p})$
\begin{equation*}
V_{k} = \prod_{j=0}^{w-1}v_{j,k}
\end{equation*}
\hrule \vspace{8pt}
Compute $p^* = (p^*_0, \cdots, p^*_k, \cdots, p^*_{n_p})$
\begin{align*}
p_0^* &= 1-V_{0} \\
p_k^* &= V_{k-1} - V_{k}  \qquad  \text{ for } k=1, \ldots, n_p-1 \\
p_{n_p}^* &= 1-V_{n_p-1}
\end{align*}
\hrule \vspace{8pt}
Compute the write amplification,
\begin{equation}\label{eqn:Hu_write_amp}
A_\text{Hu} = \frac{n_p}{n_p-\sum_{k=0}^{n_p} kp_k^*}
\end{equation}
\end{framed}
\caption{Equations needed to compute the Trim-modified empirical model-based algorithm for $A_\text{Hu}$ (adapted from~\cite{Hu2009,hu2010fundamental} and~\cite{frankie2012_ciit}).  $w$ is the window size, and allows for the windowed greedy garbage collection variation of greedy garbage collection.  Setting $w=\frac{t}{n_p}-r$ is needed for the standard greedy garbage collection discussed in this paper.  For an explanation of the theory behind these formulas, see~\cite{Hu2009,hu2010fundamental}.}
\label{fig:hu_model}
\end{figure}

\begin{figure}[!t]
\centering
\includegraphics[width=3.0in]{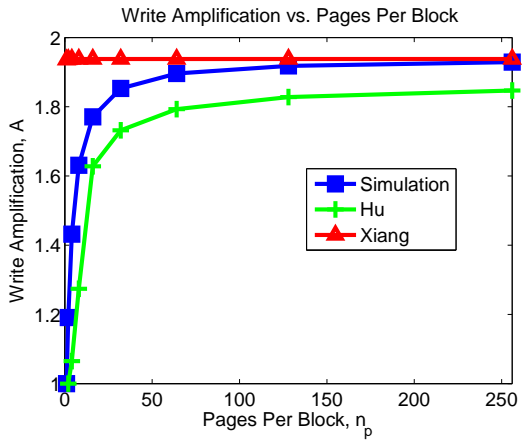}
\caption{Write amplification versus pages per block for constant effective overprovisioning.  The modified Xiang model (Equation~\ref{eqn:modified_Xiang}) fails when the number of pages per block is low.  The Trim-modified Hu model provides an optimistic, but reasonable, prediction.}
\label{fig:write_amp}
\end{figure}

Results comparing the simulated write amplification and the computed write amplification are in Table~\ref{table:write_amp} and Figure~\ref{fig:write_amp}.  In this simulation, there were 1280 blocks, with a manufacturer-specified spare factor of 0.2, and a Trim factor of 0.1.  As the number of pages per block changed, the effective spare factor was held constant to enable comparisons.  As expected, the write amplification was reduced as the number of pages per block decreased, with the minimum possible write amplification of 1 achieved with 1 page in a block.  The modification of the Xiang model shown in Equation~\ref{eqn:modified_Xiang} was a very poor approximation as the number of pages in a block decreased; this is likely due to the assumption that the number of invalid pages freed in a block by garbage collection is binomial; with a small number of pages per block, this approximation becomes too coarse.  However, the Trim-modified Hu model provides a reasonable, if optimistic, prediction of write amplification.  Note that the Trim-modified Hu model does not make sense to compute if there is one page per block, so the entry in the table is left blank.

\begin{table}
\centering
\caption{Write amplification for varying number of pages per block, equivalent to fixed sized objects that evenly divide $n_p$.  The level of effective overprovisioning was kept constant for each simulation.  The amount of Trim was $q=0.1$.  All models referenced are Trim-modified; the Xiang model was additionally modified to account for $n_p$, as shown in Equation~\ref{eqn:modified_Xiang}.}
\begin{tabular}{|m{0.5in}|m{0.5in}|m{0.5in}|m{0.5in}|} \hline
 & \multicolumn{3}{{|c|}}{Write Amplification} \\ \hline
Pages /Block $n_p$ & Simulation & Modified Xiang Model & Hu Model\\ \hline
1 & 1.000 & 1.936 & -\\ \hline
2 & 1.191 & 1.937 & 1.000\\ \hline
4 & 1.432 & 1.938 & 1.065\\ \hline
8 & 1.631 & 1.938 & 1.274\\ \hline
16 & 1.771 & 1.938 & 1.628\\ \hline
32 & 1.853 & 1.938 & 1.732\\ \hline
64 & 1.896 & 1.938 & 1.793\\ \hline
128 & 1.918 & 1.938 & 1.828\\ \hline
256 & 1.929 & 1.938 & 1.847\\
\hline\end{tabular}
\label{table:write_amp}
\end{table}

When objects do not fit neatly within blocks, such as the case for $mod(n_p, b) \ne 0$, for variable sized objects, or for objects larger than one block, the computation of write amplification is not as straightforward.  First, design decisions must be made: should an object be split over multiple blocks, or should empty, unwritten pages be left on a block, that perhaps could be filled in later by a smaller object?  The design decision will affect the analysis of the write amplification; this is an area for future research.

\section{Conclusion}
In this paper, the Trim model we previously presented in~\cite{frankie2012_acmse,frankie2012_ciit,frankie2012TOS} was extended to encompass the idea of object-based storage and allow objects to utilize more than one page for data storage.  A utilization model was derived and shown to be accurate via simulation.  Additionally, models for write amplification were modified and compared to simulated values, demonstrating the failure point of one model and the need for a more accurate model in another.  Models such as the ones presented in this paper are useful tools for understanding the problems that an FTL design needs to address, and can lead to improved future designs.

The model extension presented in this paper helps make the workload a more characteristic of real-world workloads.  By using this model in conjunction with object-based storage, we envision metadata servers that separate object classes of distinct sizes to unique flash SSDs in order to assist in self-optimization of hardware that would improve overall data access performance and improve the lifetime of the SSD.

\ifCLASSOPTIONcaptionsoff
  \newpage
\fi



\bibliographystyle{IEEEtran}
\bibliography{collectedReferences}
\end{document}